\documentclass[reprint,amsmath,amssymb,aps,superscriptaddress,prstab]{revtex4-1}
\usepackage{graphicx}
\usepackage{dcolumn}
\usepackage{bm}
\usepackage{siunitx}
\usepackage{color}
\usepackage{amsmath}
\usepackage{url} 
\usepackage{changes}
\usepackage{enumerate}
\usepackage{scalerel}

\usepackage[colorlinks, citecolor=blue, urlcolor=blue, bookmarks=false, linkcolor=blue, hypertexnames=true]{hyperref}

\begin{document}

\preprint{APS/123-QED}

\title{Comparative analysis of methods for calculating Hubbard parameters using cRPA}


\author{Indukuru Ramesh Reddy}
\affiliation{Department of Physics, Kyungpook National University, Daegu 41566, Republic of Korea}

\author{M. Kaltak} \email{merzuk.kaltak@vasp.at} 
\affiliation{VASP Software GmbH, Berggasse 21/14, 1090 Vienna, Austria}

\author{Bongjae Kim} \email{bongjae@knu.ac.kr}
\affiliation{Department of Physics, Kyungpook National University, Daegu 41566, Republic of Korea}

\date{\today}
\begin{abstract}
In this study, we present a systematic comparison of various approaches within the constrained random-phase approximation (cRPA) for calculating the Coulomb interaction parameter $U$. While defining the correlated space is straightforward for disentangled bands, the situation is more complex for entangled bands, where different projection schemes from hybridized bands to the target space can yield varying sizes of interaction parameters. We systematically evaluated different methods for calculating the polarizability functions within the correlated space. Furthermore, we analyze how different definitions of the correlated space, often constructed through Wannierization from Kohn-Sham orbitals, defines the orbital localization and play a crucial role in determining the interaction parameter. To illustrate these effects, we consider two sets of representative correlated $d$-orbital oxides: Li$M$O$_{2}$ ($M$ = V–Ni) as examples of isolated $d$-electron systems and Sr$M$O$_{3}$ ($M$ = Mn, Fe, and Co) as cases of entangled $d$-electron systems. Through this systematic comparison, we provide a detailed analysis of different cRPA methodologies for computing the Hubbard parameters.
\end{abstract}

\maketitle

\section{Introduction}

The density functional theory (DFT) is one of the most widely employed theoretical approaches for providing a realistic description of materials. However, an important shortcoming of this method lies in its treatment of correlated systems, where the innate mean-field approach of DFT is not capable to accurately capture correlated electron motions.

For a better description of the strongly correlated systems without losing the material specificity, the incorporation of the many-body Hubbard model into DFT, using both static and dynamical treatment of the Hubbard parameter $U$, has been implemented within the framework of DFT+$U$ \cite{liechtenstein1995dftu,dudarev1998dftu} and DFT+DMFT \cite{kotliar2006dftdmft,held2008dftdmft}, and produced great success in various applications \cite{kunevs2009dftdmft,kulik2015dftu,lany2008dftu,himmetoglu2014dftu}. The accuracy of these methods critically depends on the precise quantification of the interaction parameter, $U$. Several first-principles approaches have been proposed to determine parameter-free interaction parameters, including the constrained local density approximation (cLDA) \cite{dederichs1984ground,mcmahan1988calculated,hybertsen1989cLDA,anisimov1991cLDA}, linear response \cite{cococcioni2005linear,timrov2022hp},constrained random phase approximation (cRPA) \cite{aryasetiawan2004frequency,aryasetiawan2006cRPA}, pseudohybrid Hubbard density functional \cite{agapito2015acbn}, and a hybrid method between cLDA and cRPA \cite{solovyev2005hybrid}.

Among these methods, cRPA has been actively employed in the correlated electron community due to its close connection to Hubbard model approaches. The $U$ parameter in cRPA is obtained for a well-defined set of correlated bands. Key features of cRPA include the ability to select a low-energy subspace and identify individual interaction matrix elements (on-site, inter-site, interorbital, intra-orbital, and exchange) \cite{sante2023electronic,reddy2024exploring,panda2017notation}; the ability to analyze individual screening channels~\cite{vaugier2012hubbard}; and its formulation to calculate frequency-dependent $U(\omega)$ \cite{miyake2009disentaglement,sasioglu2011weighted,petochi2020u(w),werner2012satellites,neroni2019u(w),roekeghem2014u(w)}. Such features make cRPA particularly suitable for DFT extensions such as DFT+DMFT \cite{biermann2014dynamical,werner2016dynamical}. 

Despite the advantages of the cRPA approach, a few considerations must be addressed. In the cRPA procedure, one starts with a DFT calculation, and the Kohn–Sham bands are projected onto the correlated space, which then serves as the basis for computing the Coulomb interaction parameters. Two key issues in this process are the \textit{projection} procedure itself and the definition of the \textit{correlated bands}. As one starts from the hybridized Kohn-Sham orbitals, projection into a few bands, or localized basis, is required. A typical practice is the projection of the Kohn–Sham wavefunctions onto Wannier functions (WFs), a Bloch sum of orbitals \cite{wannierprojection2009aichhorn}, mostly conveniently the maximally localized Wannier functions (MLWFs) \cite{souza2001maximally,marzari2012maximally}. However, there are multiple ways to perform these projections, and they have not been systematically compared. This oversight is significant because different projection schemes yield different local basis, resulting in different Hubbard interaction parameters, hence the electronic structures of a system \cite{miyake2008screened,wannierprojectio2021karpdmft}.

Another issue is the definition of the correlated space. In practice, constructing WFs for the correlated bands is not straightforward, as these bands are typically defined by their proximity to the Fermi level. The choice of the energy window used to define the correlated space and the subsequent construction of MLWFs is not unique, especially for hybridized bands. For example, in transition metal oxide materials, $d$-bands are generally considered as the correlated bands; however, the bands near the Fermi energy often have significant contributions from ligand O-$p$ orbitals, resulting in various approaches to constructing MLWFs for the system \cite{kunes2011wannier,crpacadeo32024merkel}.


High-density $k$-point sampling enhances the accuracy of cRPA calculations but simultaneously increases computational demands. To accelerate $k$-point convergence, one might consider employing the derivative of the wavefunction; however, the validity of this approach has not been thoroughly investigated.

In this work, we systematically compare various methodologies within the cRPA framework for calculating the Hubbard parameter $U$. Our analysis focuses on the impact of different projection schemes and the correlated space definitions on the calculated interaction parameters. We demonstrate that different ways of defining the local orbitals can play a crucial role in determining the localization of correlated orbitals, particularly in systems with hybridized bands. To illustrate these effects, we investigate two representative families of $d$-orbital oxides: Li$M$O$_{2}$ ($M$ = V–Ni) as prototypical isolated $d$-electron systems and Sr$M$O$_{3}$ ($M$ = Mn, Fe, and Co) as examples of entangled $d$-electron systems. Our comprehensive study provides detailed insights into the optimal selection of cRPA methodologies to accurately determine the interaction parameters in strongly correlated materials.
 
\section{Theoretical framework}

\subsection{Brief review on cRPA}
The starting point in cRPA is the choice of an effective low-energy Hilbert space, i.e., the target space, such as the $d$-orbitals in typical transition metal complexes. This target space is usually defined as the correlated space. The method then excludes the screening channels of the correlated space from the total polarizability function $P$. The rest space polarizability ($P^{r}$) is defined as $P^{r} = P - P^{C}$ , where $P^{C}$ is the polarizability contribution from the screening within correlated space. The partially screened Coulomb interaction kernel $U$ is then calculated by solving the Bethe-Salpeter equation: $U^{-1} = [U^{bare}]^{-1} - P^{r}$, where $U^{bare}$ is the unscreened Coulomb interaction kernel. And the Coulomb interaction matrix elements among the obtained Wannier orbitals are then evaluated using the following equation:
\begin{equation}
\begin{split}
    U_{ijkl}(\omega) = \iint \mathrm{d}\mathbf{r}\mathrm{d}\mathbf{r}' \mathcal{W}_{i}^{*}(\textbf{r})\mathcal{W}_{k}^{*}(\textbf{r}')\\
    U(\textbf{r},\textbf{r}',\omega)\mathcal{W}_{j}(\textbf{r})\mathcal{W}_{l}(\textbf{r}')
\end{split}
\end{equation}
where $\omega$ is the frequency and $\mathcal{W}(\textbf{r})$ represent the correlated orbitals, which are the MLWFs.\\

Although cRPA is a general approach, its application to materials is often complicated by issue of defining the correlated orbitals, which are employed to obtain the polarizability of the correlated space, $P^{C}$. The $d$ bands that span the correlated subspace often mix with $s$-states \cite{souza2001maximally}, and in some compounds, extended $p$- states of ligands, making it difficult to uniquely identify the $d$-$d$ transitions to calculate $P^{C}$. A typical approach involves constructing the correlated subspace using Wannier functions derived from DFT Kohn-Sham orbitals \cite{marzari2012maximally,ozaki2024closest}, though this is not very transparent as there are various ways to define the Wannier functions. Several methods have been proposed to address this issue. Here, we provide a brief overview of the different approaches to calculate $P^{C}$.

\subsubsection{Band method} Calculating the polarizability of the correlated space ($P^{C}$) is relatively straightforward if the target space is well isolated from the rest space by using the Adler and Wiser equation \cite{adler1962quantum,wiser1963dielectric,aryasetiawan2004frequency}:
\begin{equation}
\begin{split}
    P^{C}(\textbf{r},\textbf{r}';\omega) = \sum\limits_{m,m'\in C}\psi_{m}(\textbf{r})\psi_{m'}^{*}(\textbf{r})\psi_{m}^{*}(\textbf{r}')\psi_{m'}(\textbf{r}')\times\\
    \left(\frac{1}{\omega+(E_{m}-E_{m'})+i\eta} - \frac{1}{\omega-(E_{m}-E_{m'})-i\eta}\right)
\end{split}
\end{equation}
where $\psi_{m}$ and $E_{m}$ are the Kohn-Sham wavefunctions and eigenvalues, respectively.\\

In realistic systems, applying this band method is often inadequate for two main reasons: $i$. The correlated subspace is not well isolated but is entangled with the rest space, necessitating disentanglement of the correlated target space. $ii$. Although the bands appear isolated, they still retain characteristics from other orbitals due to the hybridization. For example, the orbital-resolved band structure of SrCrO$_{3}$, as shown in Fig. S1 in the Supplementary Material \cite{SM}, illustrates how O-$p$ orbitals retain some character from the Cr-$d$ manifold. The following approaches have been proposed to effectively disentangle the correlated target space from the rest space to calculate $P^{C}$.\\

\subsubsection{Disentanglement method} This method involves decoupling the hybridization between the rest and correlated spaces by diagonalizing the Hamiltonian independently in each subspace \cite{miyake2009disentaglement}. The Hamiltonian for the correlated space is derived from the Wannier basis constructed within a specific energy window, resulting in two sets of fully disentangled bands. The polarizability functions $P$ and $P^{C}$ are then defined with respect to these disentangled bands, rather than the original Kohn-Sham bands.
\begin{equation}
\begin{split}
    \tilde{P}^{C}(\textbf{r},\textbf{r}';\omega) = \sum\limits_{m,m'\in C}\tilde{\psi}_{m}(\textbf{r})\tilde{\psi}_{m'}^{*}(\textbf{r})\tilde{\psi}_{m}^{*}(\textbf{r}')\tilde{\psi}_{m'}(\textbf{r}')\times\\
    \left(\frac{1}{\omega+(\tilde{E}_{m}-\tilde{E}_{m'})+i\eta}- \frac{1}{\omega-(\tilde{E}_{m}-\tilde{E}_{m'})-i\eta}\right)
\end{split}
\end{equation}\\

Here, $\Tilde{\psi}_{m}$ and $\Tilde{E}_{m}$ are the new set of wavefunctions and eigenvalues from the disentangled band structure, which differ from the $\psi_{m}$ and $E_{m}$ based on the strength of the hybridization. The correlated space, defined from the basis of Wannier functions, is strongly dependent on the chosen energy range, which in turn influences the Coulomb interaction parameters. Here, $P^{r}$ is defined from the rest space, which is orthogonal to the correlated subspace constructed form the Wannier functions.

\subsubsection{Weighted method} The weighted method reduces the dependency of $P^{C}$ on the energy window. It directly defines $P^{C}$ from the correlated subspace spanned by a set of Wannier functions. This approach introduces probabilistic weights based on projecting Bloch functions onto Wannier functions associated with correlated bands \cite{sasioglu2011weighted}. These weights represent the probability of an electron being in the correlated space both before and after the transition between electronic states, accounting for the correlated contribution of each Bloch pair. By incorporating these weights into the polarizability function, $P^{C}$ is defined without altering the Kohn-Sham band structure.
\begin{equation}
\begin{split}
    P^{C}(\textbf{r},\textbf{r}';\omega) = \sum\limits_{m,m'}p_{m}p_{m'}\psi_{m}(\textbf{r})\psi_{m'}^{*}(\textbf{r})\psi_{m}^{*}(\textbf{r}')\psi_{m'}(\textbf{r}')\times\\
    \left(\frac{1}{\omega+(E_{m}-E_{m'})+i\eta} - \frac{1}{\omega-(E_{m}-E_{m'})-i\eta}\right)
\end{split}
\end{equation}
where
\begin{center}
$p_{m} = \sum\limits_{i\in C} |T_{im}|^{2}$, $0 \leq p_{m} \leq 1$\\
\end{center}
represents the probability of an electron in $\psi_m$ being in the correlated subspace and $T_{im}$ the matrix elements of the unitary transformation matrix between a Kohn-Sham wavefunction $\psi_m$ and a Wannier function $\mathcal{W}_i$. This method neglects screening effects within the target space, as shown in the projector method. Here, the projection onto correlated orbitals is done from the Bloch functions to the Wannier functions, expressed as $|\mathcal{W}_i\rangle = \sum\limits_{m} T_{im}|\psi_m\rangle$.\\

\subsubsection{Projector~method} In the projector method, $P^{C}$ is effectively obtained by separating out the correlated target space contributions to each Bloch state \cite{kaltak2015merging}.

\begin{equation}
\begin{split}
         P^{C}(\textbf{r},\textbf{r}';\omega) = \sum\limits_{m,m'}\bar\psi_{m}(\textbf{r})\bar\psi_{m'}^{*}(\textbf{r})\bar\psi_{m}^{*}(\textbf{r}')\bar\psi_{m'}(\textbf{r}')\times\\
    \left(\frac{1}{\omega+(E_{m}-E_{m'})+i\eta} -\frac{1}{\omega-(E_{m}-E_{m'})-i\eta}\right)
\end{split}
\end{equation}

Here, $\bar\psi_m$ is the correlated Bloch function projected onto each Bloch state $\psi_n$, i.e., $|\bar\psi_m\rangle = \sum\limits_{n}P_{nm}|\psi_{n}\rangle$. The projection matrix with the correlated projector $P_{nm} = \sum\limits_{i\in C} T_{in}^{*}T_{im}$, where $i$ indexes the states in the correlated space $C$, and $T$ is the unitary transformation matrix between the Bloch and Wannier functions.

\subsection{Calculation details}
All DFT calculations were performed using the projector augmented wave (PAW) method, as implemented in the Vienna Ab initio Simulation Package (VASP) \cite{kresse1996efficient,bloch1994paw}. The exchange-correlation functional was treated using the generalized gradient approximation (GGA) with the Perdew-Burke-Ernzerhof (PBE) parametrization \cite{perdew1996gga}. The convergence criterion for self-consistent electronic energy minimization was set to 10$^{-8}$ eV. Plane wave cutoffs of 500 eV and 650 eV were used for Sr$M$O${3}$ and Li$M$O${2}$ structures, respectively. The $k$-point mesh was sampled using the following grids: $8\times8\times8$ for Sr$M$O$_{3}$, $12\times12\times2$ for LiVO$_{2}$, $6\times13\times6$ for LiMnO$_{2}$, $7\times12\times7$ for LiNiO$_{2}$, and $13\times13\times2$ for LiCrO$_{2}$, LiFeO$_{2}$, and LiCoO$_{2}$.

All structures have been considered in the nonmagnetic phase. The Wannier functions were obtained using WANNIER90 \cite{mostofi2008wannier90} and the VASP2WANNIER interface \cite{franchini2012maximally}. Default parameters were used for both the disentanglement convergence and Wannierization. Specifically, the disentanglement convergence criterion was set to a relative change of less than 10$^{-10}$ $\text{\AA}$ in the gauge-invariant part of the spread. For Wannierization, the criterion was set to an absolute change of less than 10$^{-10}$ $\text{\AA}^{2}$ in the total spread. Unless specified otherwise, we employ the $d$-$dp$ model, where the WFs are constructed within the $d$-$p$ energy window, and the screened interaction parameters are calculated specifically for the $d$-manifold in the static limit ($\omega=0$).
All interaction parameters are in the Kanamori form.

\section{Results}
\subsection{Applicability of $k-p$ perturbation theory}

\begin{figure}[h]
    \begin{center}
    \includegraphics[clip=true,scale=0.5]{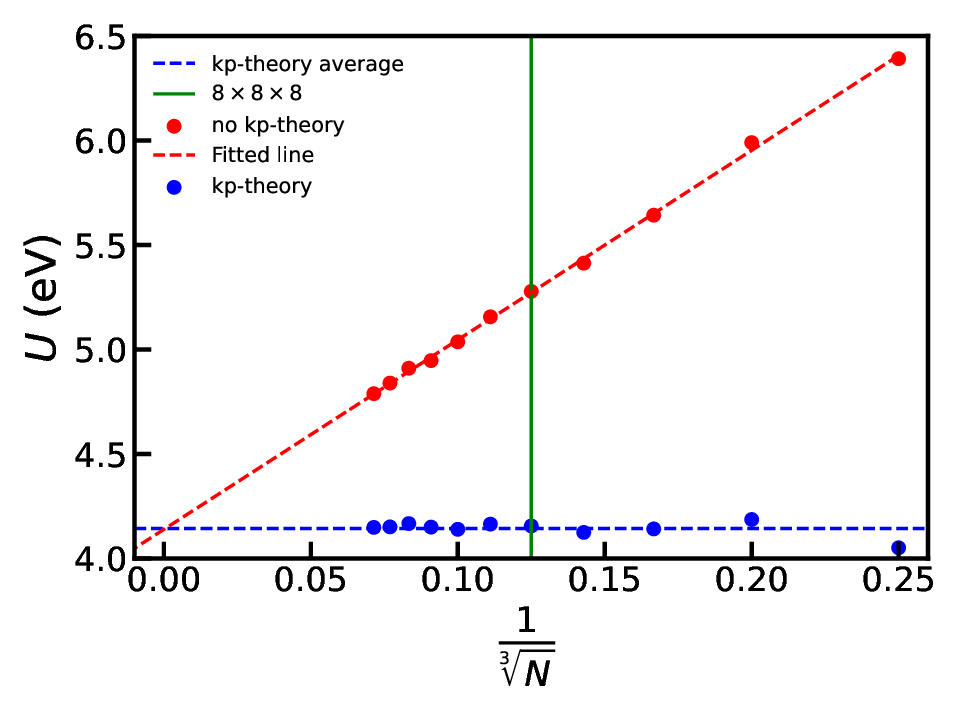}
    \end{center}
    \caption{The calculated screened Coulomb interaction parameter ($U$) of SrCrO$_{3}$ as a function of inverse of $k$-points density. $N$ is the total number of $k$-points in the Brillouin Zone. The red (blue) colors represent the $U$ values calculated without (with) $k$-$p$ perturbation theory. Extrapolation of $U$ values obtained without $k$-$p$ theory matches the converged $U$ value as $k$-points approach infinity. The vertical solid green line represents $8\times8\times8$ $k$-points.}
    \label{fig:cr-kpt}
\end{figure}

We first examine the validity of $k-p$ perturbation theory, which is guaranteed only for systems with a finite band gap \cite{AmbroschDraxl20061}. As an example, we chose SrCrO$_{3}$, a cubic perovskite transition metal oxide with well-separated $d$-bands from O-$p$ bands (see Fig. S1 in the Supplementary Material \cite{SM}). We performed cRPA calculations employing band method with  $d$-$dp$ model. The correlated bands of the SrCrO$_{3}$ is mostly of Cr-3$d$ orbital character, and the corresponding Bloch states, which are partially occupied, cross the Fermi level. In the constrained polarizability calculation, the transition between bands has finite values, and there is no zero-energy transition. As a result, one can safely evaluate the long-wave limit ${\bf q}=0$ using $k-p$ perturbation theory \cite{Gajdos2006,AmbroschDraxl20061,waveder}. In this case, $k$-point convergence is very fast, as shown in Fig.~\ref{fig:cr-kpt}, a $4\times4\times4$ $k$-point mesh is already sufficient for SrCrO$_{3}$.

In our specific example of SrCrO$_{3}$, the partially occupied Bloch bands form a correlated space. So there is no metallic contribution and the system is effectively treated as an insulator in cRPA. However, in general, states arising from entanglement and/or hybridization contribute to cRPA polarizability with zero-energy transition values. Thus, application of $k-p$ perturbation theory must be considered carefully.

For this purpose, we have calculated the screened interaction $U$ for a series of $k$-point samplings for both cases, including and excluding the long-wave limit. It is known that the $k$-point integration error of the RPA energy decays with $1/N_k^2$ for metals \cite{Harl2010}. Thus, it can be expected that the same error in the screened interaction $U$ decays with $1/N_k$. This relationship is demonstrated in Fig.~\ref{fig:cr-kpt}. The difference in $U$ values, calculated with and without the long-wave limit in polarizability, decreases linearly as the number of $k$-points increases.
Even with a larger number of $k$-points of $14\times14\times 14$, the value is slightly off by 13\% for this case.

One note here is that in the band method, isolated $d$-bands are treated without considering the hybridization of other orbitals, such as O-$p$ bands, during the screening process. But in fact, while there is clear energy separation between the isolated $d$-bands and the rest space, there still exists a sizable hybridization, and related projections are not properly taken into account in the band method. As a result, weaker screening results in larger Hubbard $U$ values. On the other hand, the projector method with stronger screening yields reduced $U$ values compared to the band method. For the common $k$-points density, where the full convergence is not reached, there can be an overestimation of the Hubbard parameters. In our previous work, we adopted the projector scheme and employed $8\times8\times8$ $k$-points, which is marked with a green line in Fig.~\ref{fig:cr-kpt} \cite{reddy2024exploring}. The case for SrVO$_{3}$ is provided in the Supplementary Material \cite{SM}.

\subsection{Isolated $d$-manifold: Cases for Li$M$O$_{2}$ ($M$ = V - Ni)}

Let us first investigate the isolated band systems. Among the transition metal oxides family, we have chosen a set of layered cathode materials, Li$M$O$_{2}$, with $M$ = V, Cr, Mn, Fe, Co, and Ni (see Fig. S3 in the Supplementary Material \cite{SM}). The Li$M$O$_{2}$ series exhibit well-isolated $d$-bands, as shown in Fig.~\ref{fig:orb-resolv-lmo} and Fig. S4 in the Supplementary Material \cite{SM}, making them suitable candidates for comparing the two different schemes: (i) the band method and (ii) the projector method. Even for isolated band systems, where the target bands are energetically separated from the other bands, there still exist sizable contributions from the rest space, as shown in Fig.~\ref{fig:orb-resolv-lmo}. Hence, the band method, where the target bands themselves are considered as correlated subspace, is different from the projector method, where the correlated subspace is projected out. The other schemes previously introduced for the entangled bands will be discussed in the next section.

\begin{figure}[!htb]
    \begin{center}
    \includegraphics[clip=true,scale=0.32]{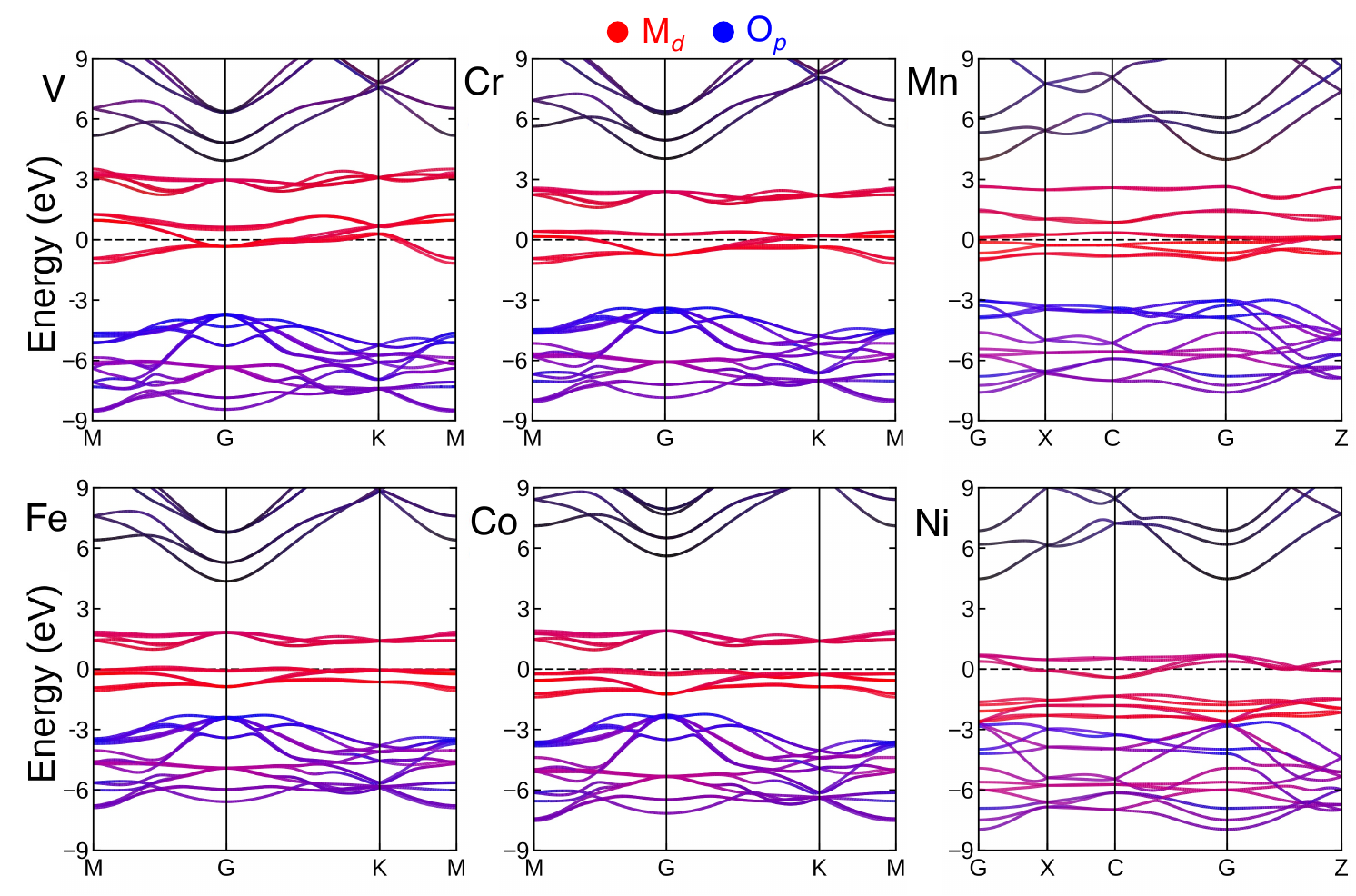}
    \end{center}
    \caption{Orbital-resolved band structures of Li$M$O$_{2}$, with TM-$d$ and O-$p$ bands projected in red and blue, respectively.}
    \label{fig:orb-resolv-lmo}
\end{figure} 

\begin{figure}[!htb]
    \begin{center}
    \includegraphics[clip=true,scale=0.47]{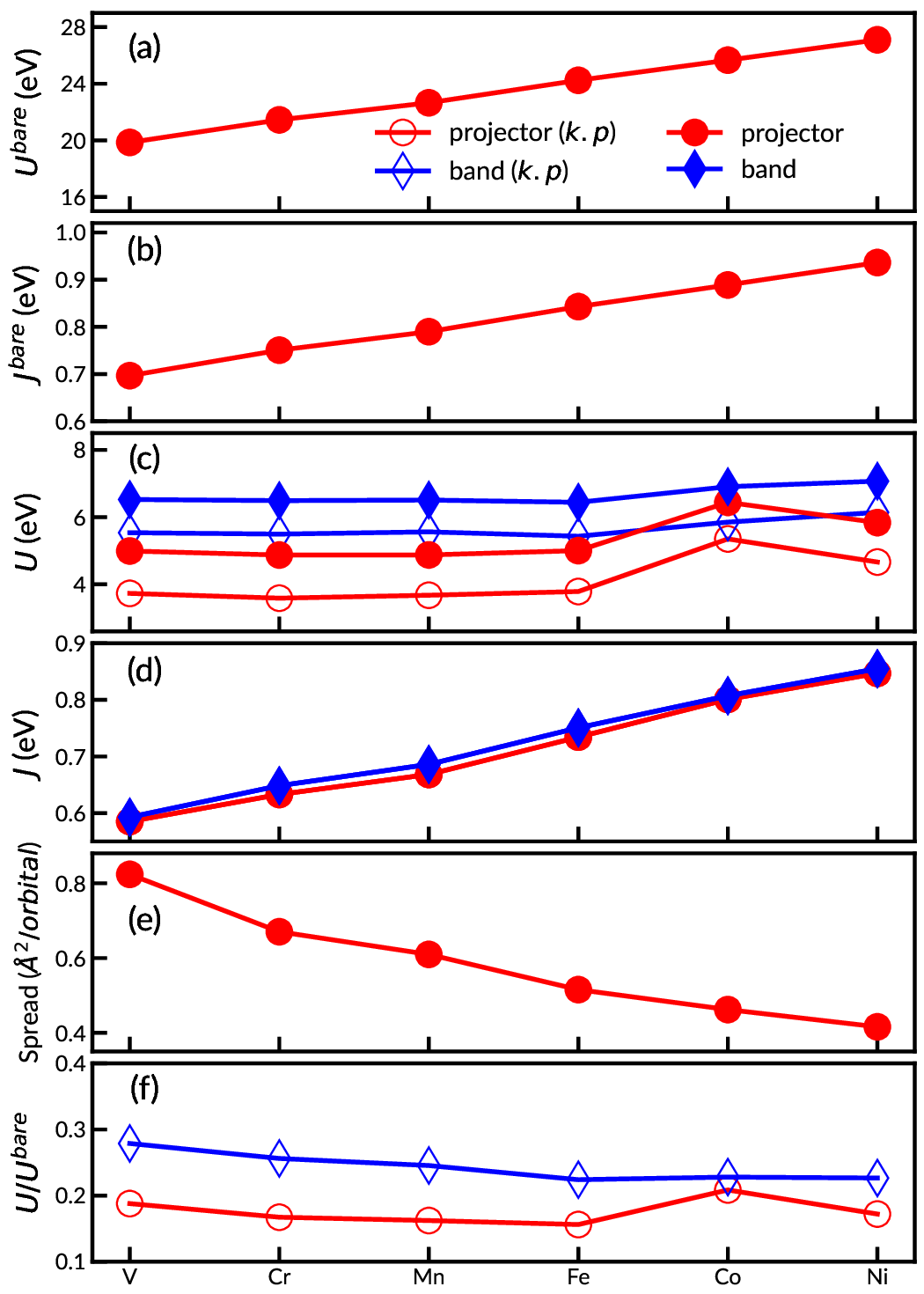}
    \end{center}
    \caption{Evolution of the cRPA Hubbard parameters ($U^{bare}$, $J^{bare}$, $U$, and $J$) and the spread of Wannier functions. The red (blue) solid lines represent the values obtained using the projector (band) method within the $d$-$dp$ model. Open and filled symbols represent the values calculated with and without $k$-$p$ perturbation theory, respectively. The unscreened Hubbard parameters ($U^{bare}$ and $J^{bare}$) and the spread of Wannier functions are the same for all methods.}
    \label{fig:crpa-lmo}
\end{figure} 

\begin{figure}[h]
    \begin{center}
    \includegraphics[clip=true,scale=0.15]{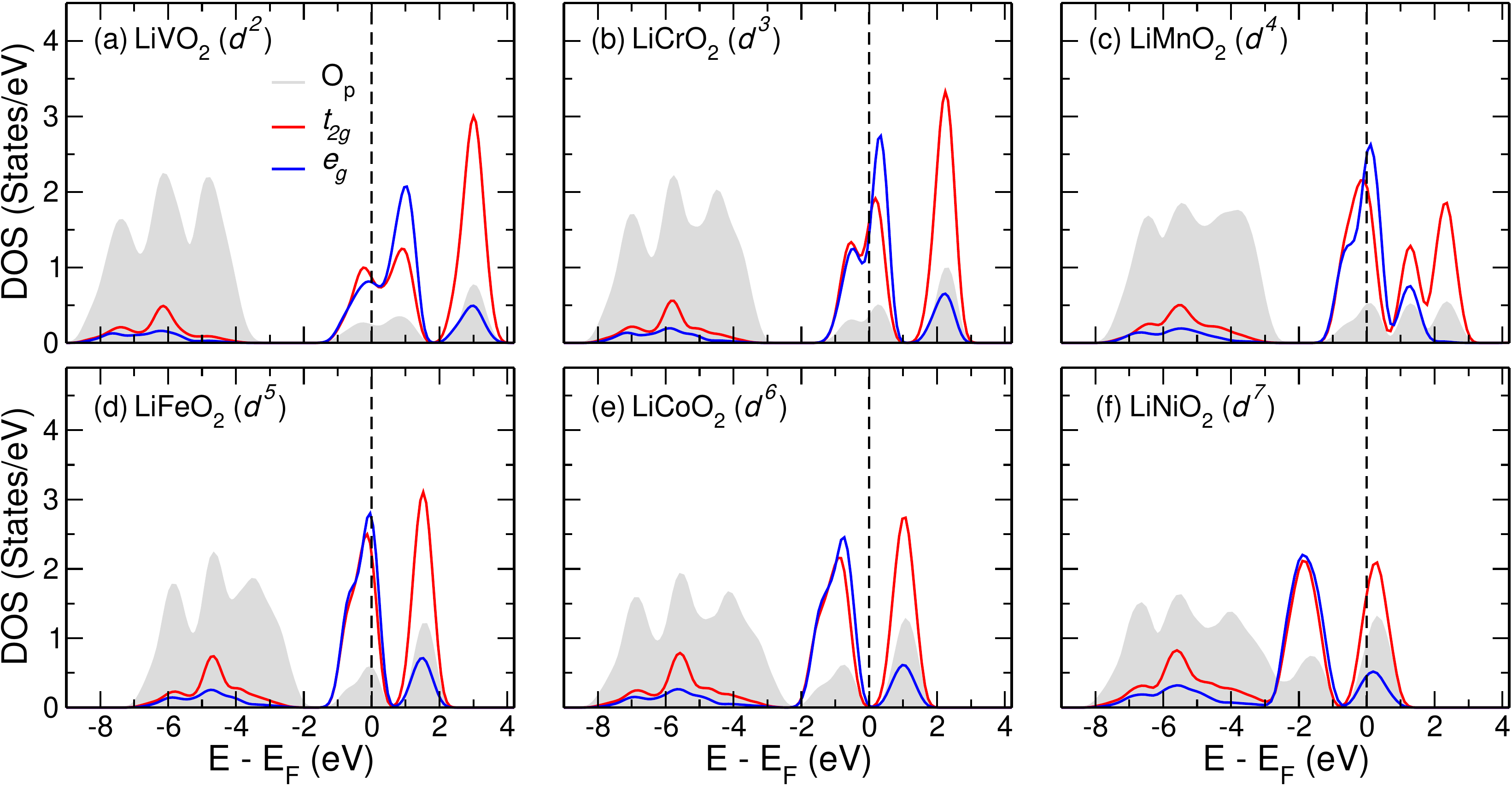}
    \end{center}
    \caption{Partial density of states (PDOS) of Li$M$O$_{2}$. The Fermi level (E$_{F}$) is set to 0 eV.}
    \label{fig:pdos-1x-lmo}
\end{figure} 

The corresponding results are shown in Fig.~\ref{fig:crpa-lmo}. For the bare Hubbard parameters ($U^{bare}$ and $J^{bare}$), no screening is included at this stage, resulting in method-independent values (Fig.~\ref{fig:crpa-lmo}(a) and (b)). However, variations arise in the screened Hubbard parameters depending on the method used (Fig.~\ref{fig:crpa-lmo}(c) and (d)).\\

First, we observe that $U^{bare}$ and $J^{bare}$ increase systematically from early to late transition metal series. This trend reflects enhanced spatial localization of $d$-orbitals with increasing occupation, as evidenced by a decrease in Wannier spread, which quantify orbital localization \cite{reddy2024exploring,vaugier2012hubbard,kim2021quantification}. However, the screened $U$ values do not strictly follow this localization trend as shown in Fig.~\ref{fig:crpa-lmo}(c) and exhibit variability of the values depending on the method (band vs. projector) and whether $k$-$p$ perturbation theory is employed to accelerate the $k$-point convergence. We used a $k$-mesh density ranging from 110 to 160 $k$-points per $\text{\AA}^{-3}$, which is in general more than enough for total energy calculations on the DFT level, but results in overestimated screened $U$ values of about 15 - 34 \% in the cRPA calculations without exploiting $k-p$ perturbation theory (compare filled and unfilled symbols). As the increase of the $k$-mesh density can be prohibitively expensive, one can resort employing the long-wave limit.\\

If one compares both methods, we clearly see that the overall screened interaction obtained using the band method are consistently larger than the one predicted by the projector method. In our isolated band cases, the differences are roughly of 1 eV. The reason for this is that in the band method, where isolated bands are directly considered as correlated subspace, the screening process from the hybridization is not considered. Due to the hybridization, even for the seemingly separated bands, the rest space has a sizable contribution from the correlated TM-$d$ orbitals, and the isolated $d$-bands also have a contribution from the rest bands, especially O-2$p$ bands in our case (see orbital-resolved band structures in Fig.~\ref{fig:orb-resolv-lmo}). As a result, effective screening is curtailed in the band method compared to the projector method, where the projection from the Bloch bands to correlated subspace are performed, and this yields larger $U$ values.

In general, the screened interaction parameters are determined from the competition between Wannier localization and screening. Both effects are enhanced upon occupation -- more charges localize the orbitals, and occupation decreases the $d$-$p$ gap, hence, increases the hybridization. The resulting evolution of interaction parameters can be different for different set of systems and may sometimes be nonmonotonic \cite{vaugier2012hubbard,kim2018strain,si2025evolution,reddy2024exploring}. As the screening is not effective for Hund interaction parameters and they follows the trends of the bare values as in Fig.~\ref{fig:crpa-lmo}. As screened $U$ values do not always increase or often decrease upon occupation~\cite{reddy2024exploring}, the role of the Hund interactions can be very effective for the later transition metal compounds \cite{haule2009coherence,ryee2021hund,georges2013strong,kang2021optical,lee2021hund,karp2020role,bramberger2021hund,springer2020hund,yin2011magnetism}.

It has been reported that underestimation of the gap (such as the band gap or the $d$-$p$ splitting) in transition metal oxides, especially when using typical (semi-)local DFT functionals, leads to overscreening in cRPA \cite{werner2016dynamical}. One potential way to address this issue is by using more accurate \textit{ab initio} methods, such as modifying the band structure with an empirical +$U$ correction on the ligand sites \cite{carta2025importance} or replacing DFT with the GW method \cite{werner2016dynamical}. 

Note that the underestimation of gaps in DFT calculations is compensated by the absence of local exciton vertex corrections at the RPA level, thus offsetting the issue of overscreening \cite{loon2021vertex}. This justification applies to the band method, where the target bands within the correlated space are excluded from the screening, treating the system as an insulator. In contrast, the projector method considers all bands (including the $d$-bands near the Fermi level) for screening by projecting out the correlated space contribution from each band. cRPA has a general tendency to overscreen states near the Fermi level (E$_\textit{F}$) \cite{Honerkamp2018limitation-crpa,Honerkamp2021overscreen}. As a result, as shown in Fig.~\ref{fig:crpa-lmo}, the projector method underestimates the $U$ values. 

The projector method exhibits nonmonotonicity in $U$ values, as evidently shown for LiCoO$_{2}$, where a significantly larger $U$ value is obtained. This is not observed in the band method. In a non-spin-polarized setup, Co$^{5+}$ in LiCoO$_2$ has a formal occupation of $t{2g}^{6}e{g}^{0}$, making the system a gapped insulator (see Fig.~\ref{fig:pdos-1x-lmo} and Fig. S4 in the Supplementary Material \cite{SM}). This can be attributed in part to the weaker screening in systems where the electronic states are farther from the E$_\textit{F}$, compared to those where the electronic states are closer to E$_{F}$. Thus, the screening appears to be incomplete in the case of LiCoO$_{2}$. Hence, this case is different from the other nonmonotonicity upon occupancies~\cite{kim2018strain,reddy2024exploring}, rather, it can be considered as an artifact, and possible extension to the spin-dependence is expected to remedy this~\cite{Linscott2018}.

\subsection{Entangled $d$-manifold: Cases for Sr$M$O$_{3}$ ($M$ = Mn, Fe, and Co)}

We now discuss the cases for the entangled bands, focusing on three transition metal perovskites, Sr$M$O$_{3}$ ($M$ = Mn, Fe, and Co), which exhibit strong hybridization between TM-$d$ and O-$p$ bands (see Fig. S5 and Fig. S6 in the Supplementary Material \cite{SM}). In cRPA calculations of the Coulomb interaction parameters, a key part is to obtain the polarizability of the correlated subspace, $P^{C}$. There are several schemes for this purpose, as described before. Here, we systematically evaluate the projector, weighted, and disentanglement methods.\\

The choice of energy window for constructing the Wannier functions, along with the screening channel, enables us to define various models, such as $t_{2g}$-$t_{2g}$, $e_{g}$-$e_{g}$, $t_{2g}$-$d$, $e_{g}$-$d$, $d$-$d$, $d$-$dp$, $dp$-$dp$, etc. This nomenclature, as described in Refs. \cite{vaugier2012hubbard,miyake2008d,panda2017notation}, uses the first index to specify the correlated subspace for which the Hubbard $U$ is calculated, and the second index to define the bands (basis) used in constructing the Wannier functions. For example, in the $d$-$d$ model, Wannier functions are constructed from the $d$ bands, and the Hubbard $U$ is calculated specifically for these $d$ bands. On the other hand, in the $d$-$dp$ model, Wannier functions are constructed using both metal $d$ and ligand $p$ bands, while the Hubbard $U$ is calculated only for the bands in the $d$ manifold. The latter models are also referred to as hybrid models.\\

The choice of how to define the correlated subspace is crucial to evaluate the Coulomb interaction parameters in cRPA calculations. Specifically, different choices of energy windows (outer and/or frozen) or bands used to construct the Wannier orbitals, obtained by projecting the DFT bands, can affect the interaction parameters. We primarily focus on two types of Wannier orbitals: frontier orbitals (FO) and atomic orbitals (AO).\\

\begin{figure}[h]
    \begin{center}
    \includegraphics[clip=true,scale=0.15]{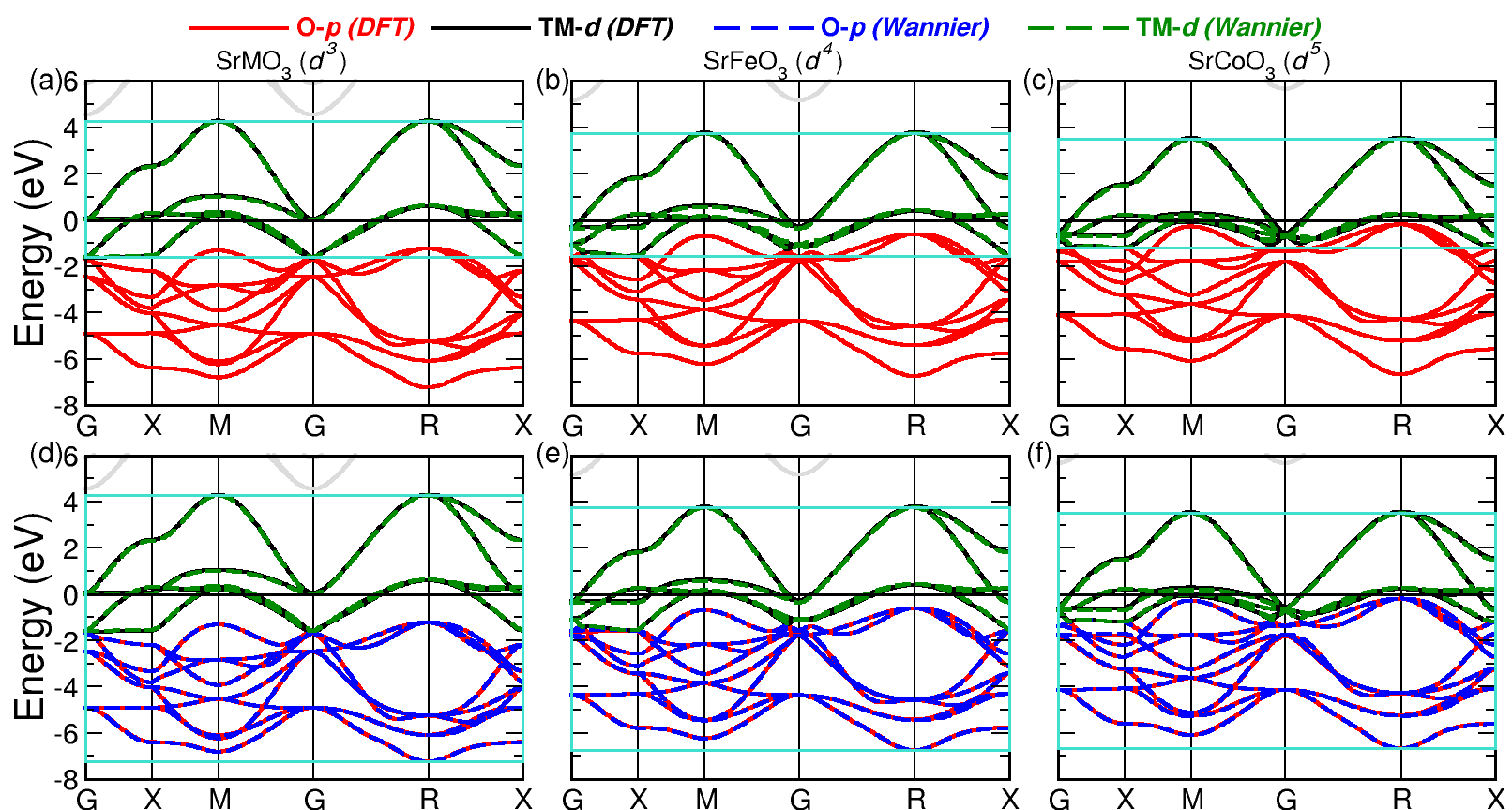}
    \end{center}
    \caption{The nonmagnetic DFT band structures, along with their Wannier-projected bands, for a simple cubic Sr$M$O$_{3}$. (a – c) FO basis (d – f) AO basis. The solid black (red) and dashed green (blue) lines represent the $Ab~initio$ and Wannier-projected bands of TM-$d$ (O-$p$), respectively. The outer energy windows for constructing the Wannier functions are represented by turquoise lines.}
    \label{fig:smo-bs}
\end{figure}

The main difference between constructing the FO and AO sets lies in the choice of the energy window and/or bands. The construction of Wannier functions using an energy window that typically covers the dominant TM-$d$ character bands (or $d$-manifold) results in the FO. Extending the energy window to include both the dominant TM-$d$ and ligand $p$ character bands results in the AO. In total, we construct five sets of Wannier orbitals using $d$-$d$ and $d$-$dp$ models, employing different approaches: three sets belong to the frontier basis / frontier orbitals (FO) and two sets to the localized basis / atomic-like orbitals (AO). For simplicity, we demonstrate one FO and one AO set here, and a more detailed description of the other choices can be found in the Supplementary Information. The range of energy windows used to construct the FO and AO are given in Table.~\ref{table:energy-windows} and also indicated in Fig.~\ref{fig:smo-bs}.\\

\begin{figure}[h]
    \begin{center}
    \includegraphics[clip=true,scale=0.6]{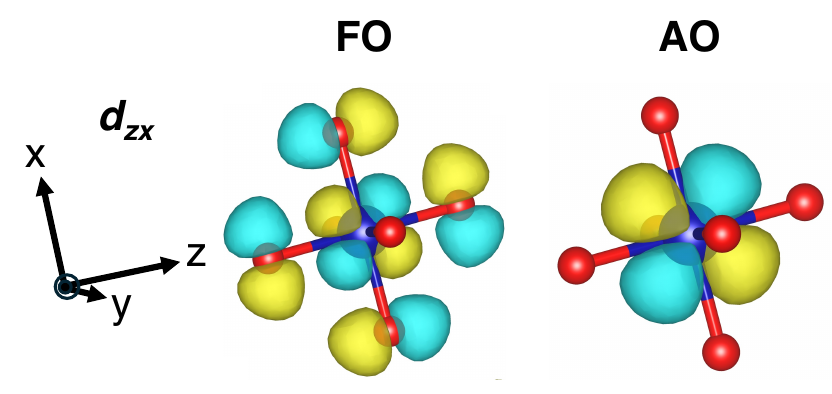}
    \end{center}
    \caption{Isosurfaces of the frontier (FO) and atomic (AO) $d_{zx}$ Wannier orbital in SrCoO$_{3}$ at the same isovalue.}
    \label{fig:xz-Co}
\end{figure}

The appearance of the FO and AO centered on the TM is straightforward to understand. The bands within the extended $dp$-energy window consist of both TM-$d$ and O-$p$ character (bonding and antibonding) due to band entanglement and hybridization. When a small outer energy window and/or a frozen energy window that includes only $d$-like bands is selected to construct the Wannier functions, the antibonding character of the underlying hybridization becomes apparent, with a node between the TM and O sites (see Fig.~\ref{fig:xz-Co}). For clarity, we show only one $d$ orbital ($d_{xy}$) for SrCoO$_{3}$ in Fig.~\ref{fig:xz-Co}, while all Wannier orbitals are shown in Fig. S8 (see Supplementary Materiel \cite{SM}).\\

\begin{table}[h]
\begin{center}
\caption{The outer energy window ${\mathbb{W}}$ (in eV) for the FO and AO basis for Sr$M$O$_{3}$ ($M$ = Mn, Fe, and Co).}
\label{table:energy-windows}
\begin{tabular*}{0.35\textwidth} {c c c}
\hline
\hline
 ~ & FO & AO \\
\hline
Mn & [-1.60,4.26] & [-7.24,4.26] \\
Fe & [-1.54,3.74] & [-6.75,3.74] \\
Co & [-1.21,3.50] & [-6.67,3.50] \\
\hline
\end{tabular*}
\end{center}
\end{table}

A larger outer energy window, encompassing both localized target TM-$d$ orbitals and the rest space, results in more localized Wannier functions resembling AO. These localized Wannier functions centered on Co do not exhibit the $p$-like tails, but they still exhibit $s$-like tails on O atoms (in the $e_{g}$ orbitals), enforced by hybridization and orthogonalization between Wannier functions centered on different sites (see Fig. S8 in the Supplemetary Material \cite{SM}). The weights on the surrounding O atoms are substantially less in the AO basis compared to the FO basis.\\

Starting from the SrMnO3 ($d^{3}$), a half-filled $t_{2g}$ band, electrons occupy the $t_{2g}$ manifold in SrFeO$_{3}$ ($d^{4}$) and SrCoO$_{3}$ ($d^{5}$). The O-$p$ bands, initially entangled with $t_{2g}$ bands with an overlap in the energy range of 0.37 eV in SrMnO$_{3}$, shift to higher energies in SrFeO$_{3}$ and SrCoO$_{3}$, becoming entangled over energy ranges of 0.93 eV and 1.01 eV, respectively, indicating enhanced $pd$-hybridization. The decreasing difference between the TM-$d$ and O-$p$ band centers with increasing electron occupation further confirms this enhanced $pd$-hybridization. Specifically, the energy differences between the TM-$d$ and O-$p$ band centers are 2.29 eV, 1.52 eV, and 0.93 eV for SrMnO$_{3}$, SrFeO$_{3}$, and SrCoO$_{3}$, respectively. Thus, As electron filling increases, contributions from the rest space (O-$p$ bands) to the correlated space become more significant. Consequently, the $d$-$p$ screening channel plays key role in the screening process. This behavior differs from previously observed cases involving isolated $d$-manifolds.\\

\begin{figure}[h]
    \begin{center}
    \includegraphics[clip=true,scale=0.38]{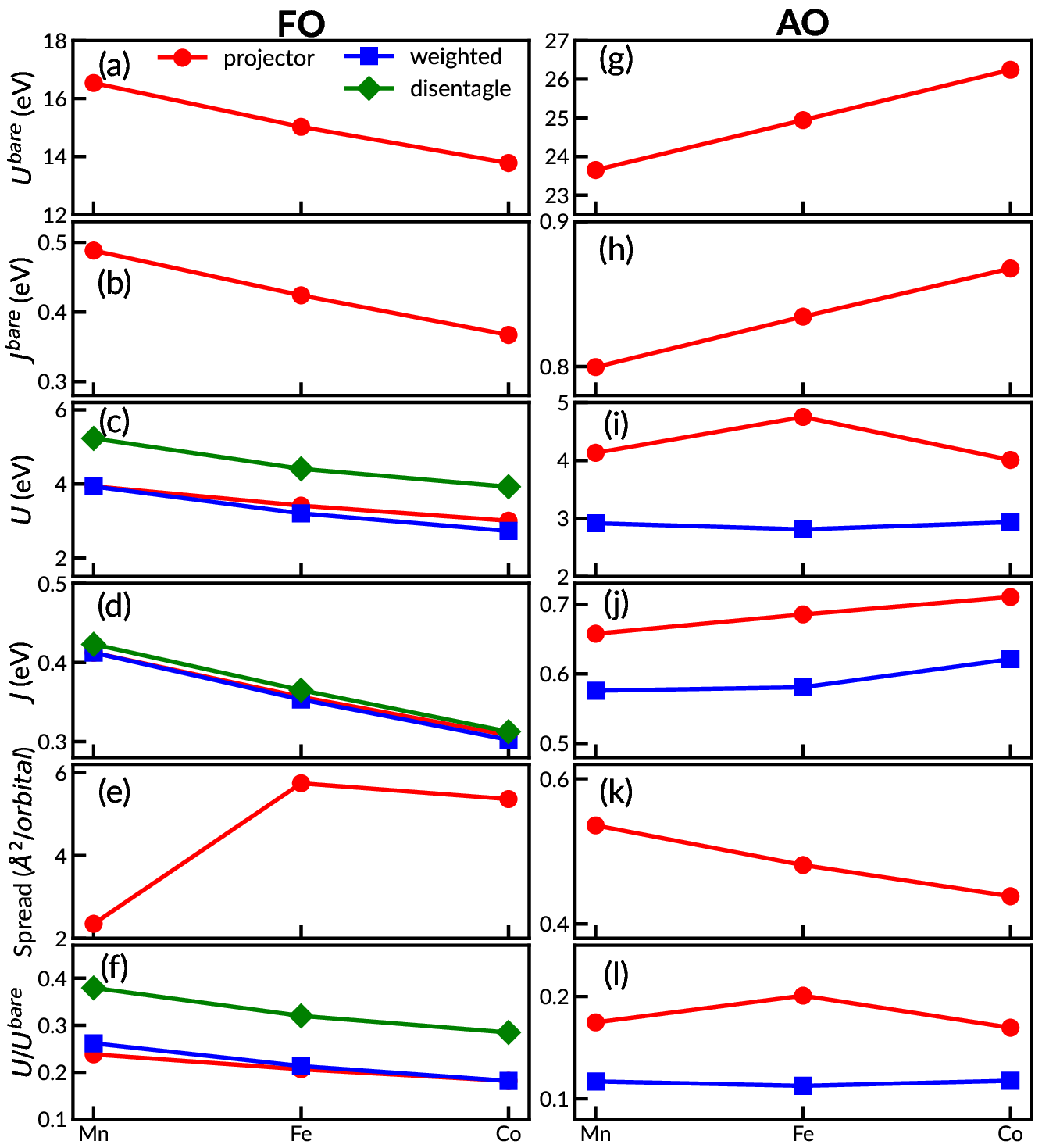}
    \end{center}
    \caption{Evolution of the cRPA Hubbard parameters ($U^{bare}$, $J^{bare}$, $U$, and $J$) and the spread of Wannier functions within (a - f) FO1 basis, and (g - l) AO2 basis. The unscreened Hubbard parameters ($U^{bare}$ and $J^{bare}$) and the spread of Wannier functions are the same for all methods.}
    \label{fig:crpa-smo}
\end{figure}

We compare interaction parameters calculated using different cRPA methods: the projector method, the weighted method, and the disentanglement method.
Here, we note that the disentanglement method always uses the minimal basis set, which is $d$-bands for the current case. This removes the distinction from the hybrid model; the basis set size is always of the same size as the correlated space. As a result, this method is limited to non-hybrid models, such as $t_{2g}$-$t_{2g}$ and $d$-$d$ models. However, disentanglement method is strongly dependent on the defined energy window, introducing a sizable degree of arbitrariness. For this reason, it has not been applied to the $d$-$dp$ model (AO basis).
First, we focus on the unscreened $U^{bare}$ values. As seen in the band structure and PDOS (Fig.~\ref{fig:smo-bs} and Fig. S5 in the Supplementary Material \cite{SM}), the $t_{2g}$ band narrows with increasing electron filling, indicating enhanced localization. However, despite the localization of $d$-orbitals, $U^{bare}$ values decrease as electron filling increases in the case of FOs (Fig.~\ref{fig:crpa-smo}). This trend is further reflected in the increased spread of the Wannier functions obtained from the frontier basis (FO) (Fig.~\ref{fig:crpa-smo}(e)) compared to the localized basis (AO) (Fig.~\ref{fig:crpa-smo}(k)). This is because, despite the strong $d$-$p$ hybridization, the Wannierization process is done in the narrow energy ranges, which leads to insufficient basis, giving larger Wannier spread. As a result, we have in much smaller $U^{bare}$ for FO and observe a decreasing trend in $U$ with occupation (Fig.~\ref{fig:crpa-smo}(a) and (c)) along with $U^{bare}$. Notably, this trend is consistent across the other cRPA methods as well (see Fig. S9 in the Supplementary Material \cite{SM}). \\

Constructing the Wannier basis from AO results in enhanced localization with increasing electron occupation, as evidenced by the increase in $U^{bare}$ and the corresponding decrease in Wannier spreads (Fig.~\ref{fig:crpa-smo}(g) and (k)). However, the screened Coulomb interaction $U$ does not follow this tendency (Fig.~\ref{fig:crpa-smo}(i)), indicating that simple Wannier localization is not the dominant factor. Instead, a competition arises between basis localization, which increases the Coulomb interaction, and enhanced screening from the $d$-$p$ channel, leading to more complex behavior.
Note the small interaction parameters reported in Ref.~\cite{si2025evolution}, where $k-p$ perturbation theory is employed.\\

The screened $U$ for the localized basis arises from far more complex physics, driven by the active involvement of the rest space (\textit{i.e.}, $d$-$p$ screening channel). Similar to the DFT+$U$ approach, where the Hubbard $U$ term is added based on atomic orbitals that resemble localized orbitals, it cannot simply be assumed that $U$ increases with electron occupation. Furthermore, strong hybridization between the screening subspace and correlated orbitals results in calculated Coulomb parameters that depend on the method employed. Different projection schemes onto the correlated subspace from entangled bands yield significantly different $U$ values (see Supplementary Material for more details \cite{SM}).\\

As disentanglement method is highly dependent on the energy window, the resulting interaction parameters are not unique, especially for the hybridized bands. As shown in Fig.~\ref{fig:crpa-smo}(c), we see that while the projector and weighted schemes are similar, the disentanglement scheme gives different $U$. For AO, from the setup, we only practice the projector and weighted method. 
Overall, the weighted method produces smaller $U$ values than the projector method because it removes fewer screening effects from the fully screened polarizability. This difference is especially pronounced in the AO basis, as illustrated in Fig.~\ref{fig:crpa-smo}(i) and (j).
Employing a correlated projector rather than the probability approach seems to correct the overscreening.\\

\section{Conclusions}
In summary, we have systematically investigated the calculation of Hubbard $U$ using various cRPA methods, providing a comprehensive analysis of their specific implementations. We explored different approaches to construct the Wannier basis for extracting the correlated subspace from the projection of DFT Kohn-Sham orbitals, while addressing the complexities inherent in defining the correlated basis. Our study highlights that both the choice of method and the construction of the Wannier basis play a key role in influencing the quantification of Hubbard $U$, providing valuable insights for future studies focused on the determination of interaction parameters in correlated materials.\\

We believe that a systematic comparison of different Hubbard interaction parameter sets within the DFT+$U$ and/or DFT+DMFT frameworks, especially in terms of physical observables, will further reveal the practicality of the diverse methods and basis sets \cite{ricca2020-UV,kim2021quantification,macke2024-U,reddy2024exploring,reddy2025-UV}. Further insight can be gained by analyzing the frequency dependence $U(\omega)$ across the different methods discussed here, as screening is fundamentally frequency dependent. A detailed study on how the basis choices influence structural and electronic properties, along with an analysis of frequency dependence $U(\omega)$, remains an important direction for future work.\\

\begin{acknowledgments}
We acknowledges support from NRF (NRF-2021R1A4A1031920, RS-2021-NR061400, and RS-2022-NR068223) and KISTI Supercomputing Center (Project No. KSC-2023-CRE-0413)
\end{acknowledgments}

\bibliography{main}

\end{document}


\preprint{APS/123-QED}

\title{Supplementary Material of \\ ``Comparative analysis of methods for calculating Hubbard parameters using cRPA"}

\author{Indukuru Ramesh Reddy}
\affiliation{Department of Physics, Kyungpook National University, Daegu 41566, Republic of Korea}

\author{M. Kaltak} \email{merzuk.kaltak@vasp.at} 
\affiliation{VASP Software GmbH, Berggasse 21/14, 1090 Vienna, Austria}

\author{Bongjae Kim} \email{bongjae@knu.ac.kr}
\affiliation{Department of Physics, Kyungpook National University, Daegu 41566, Republic of Korea}

\date{\today}

\maketitle


\onecolumngrid
\section{Supplementary Note 1}
We investigate the $k$-point convergence of the screened Coulomb interactions for SrVO$_{3}$ using the $d$-$dp$ model. With a dense $k$-point grid of $14\times14\times14$, the $U$ value deviates by 12.6\% from the converged $U$ value. Increasing the $k$-point density towards infinity leads to the converged $U$ value, as seen in the case of SrCrO$_{3}$ (\textbf{Fig. 1} in the main text). The vertical green line in Fig.~\ref{fig:v-kpt} indicates the number of $k$-points ($8\times8\times8$) used in our previous study \cite{reddy2024exploring}, where the Hubbard $U$ value, calculated using the projector method, was approximately 5.7\% higher than the converged $U$ value obtained from band method. Unlike SrCrO$_{3}$, the $d$-bands of SrVO$_{3}$ are not well isolated, exhibiting slight entanglement between the V-$e_{g}$ bands and higher unoccupied bands (Fig.~\ref{fig:orb-bs-v-cr}). Thus, we further calculated the $U$ value within the same framework, but using $t_{2g}$-$t_{2g}$$p$ model, which does not include the entangled $e_{g}$ bands. Interestingly, in this case as well, the $U$ value of the projector method, using a $8\times8\times8$ $k$-point grid, is approximately 6\% higher than the converged $U$ value of the band method (data not shown here).\\

\renewcommand{\figurename}{Fig.}
\renewcommand{\thefigure}{S\arabic{figure}}
\begin{figure}[h]
\begin{center}
\includegraphics[clip=true,scale=0.5]{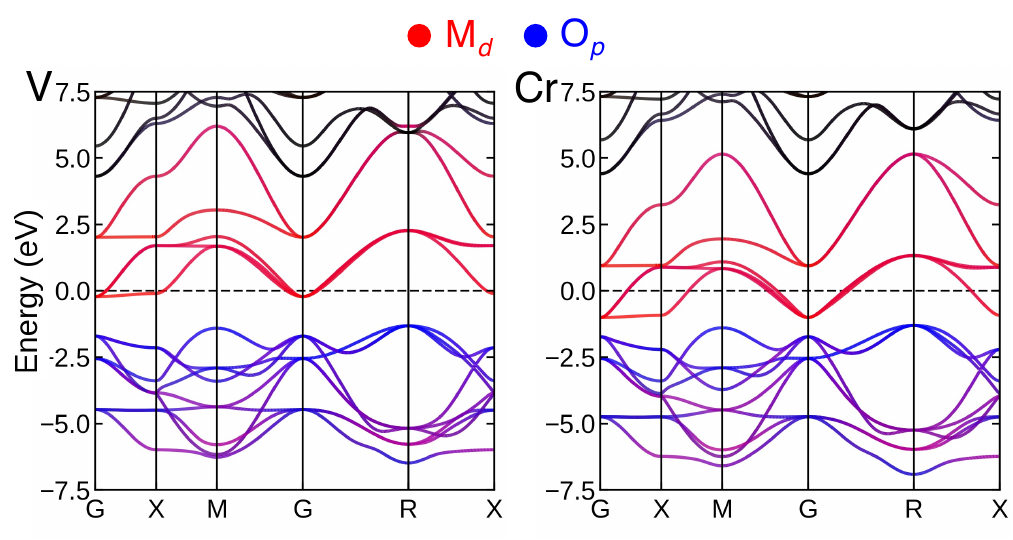}
\end{center}
\caption{Orbital-resolved band structures of Sr$M$O$_{3}$ ($M$ = V and Cr), with TM-$d$ and O-$p$ bands are projected in red and blue, respectively.}
\label{fig:orb-bs-v-cr}
\end{figure} 

\renewcommand{\figurename}{Fig.}
\renewcommand{\thefigure}{S\arabic{figure}}
\begin{figure}[!htb]
\begin{center}
\includegraphics[clip=true,scale=0.5]{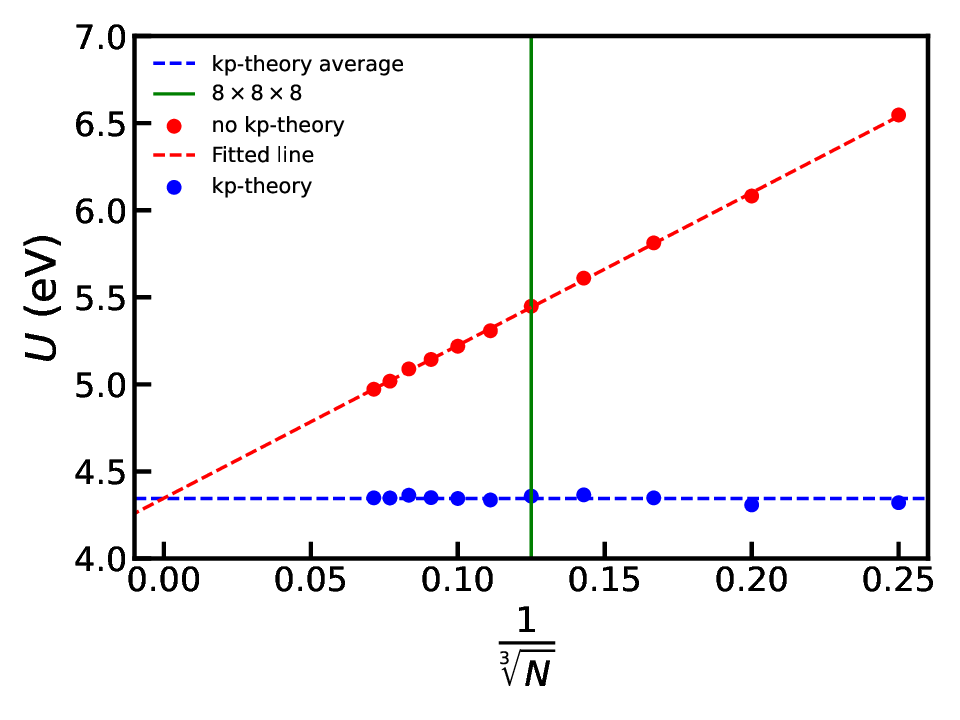}
\end{center}
\caption{The calculated screened Coulomb interaction parameter ($U$) of SrVO$_{3}$ as a function of $k$-points numbers. The red (blue) colors represent the $U$ values calculated without (with) long-wave limit of polarizability. The vertical solid green line represents the $8\times8\times8$ $k$-points.}
\label{fig:v-kpt}
\end{figure} 

\section{Supplementary Note 2}
The Li$M$O$_{2}$ structures considered for the cases of isolated $d$-band materials crystallize in the hexagonal R$\bar{3}$m structure ($M$ = V, Cr, Fe, Co) and the monoclinic C2/m structure ($M$ = Mn, Ni), as shown in Fig.~\ref{fig:lmo-structures}. The corresponding band structures, along with the Wannier projected bands, are shown in Fig.~\ref{fig:lmo-bs-1x}.\\

\renewcommand{\figurename}{Fig.}
\renewcommand{\thefigure}{S\arabic{figure}}
\begin{figure}[h]
\begin{center}
\includegraphics[clip=true,scale=0.5]{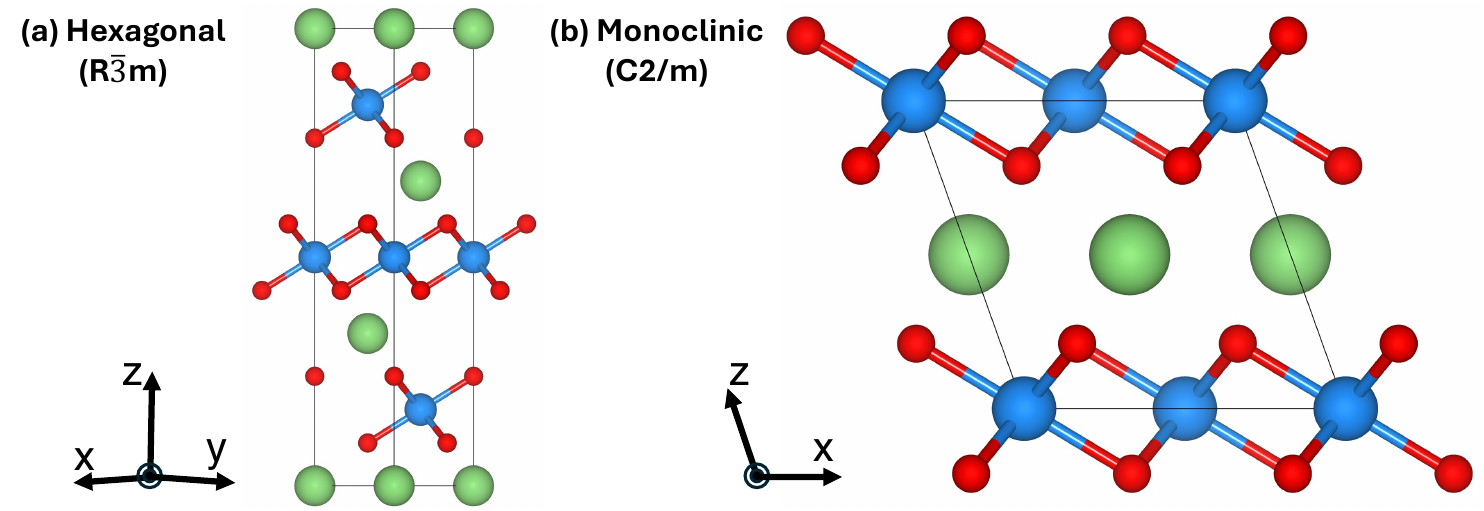}
\end{center}
\caption{Crystal structures of (a) Li$M$O$_{2}$ ($M$ = V, Cr, Fe, Co) in the R$\bar{3}$m space group, and (b) Li$M$O$_{2}$ ($M$ = Mn, Ni) in the C2/m space group. The green, sky-blue, and red spheres indicate Li, $M$, and O atoms, respectively.}
\label{fig:lmo-structures}
\end{figure} 

\renewcommand{\figurename}{Fig.}
\renewcommand{\thefigure}{S\arabic{figure}}
\begin{figure}[h]
\begin{center}
\includegraphics[clip=true,scale=0.26]{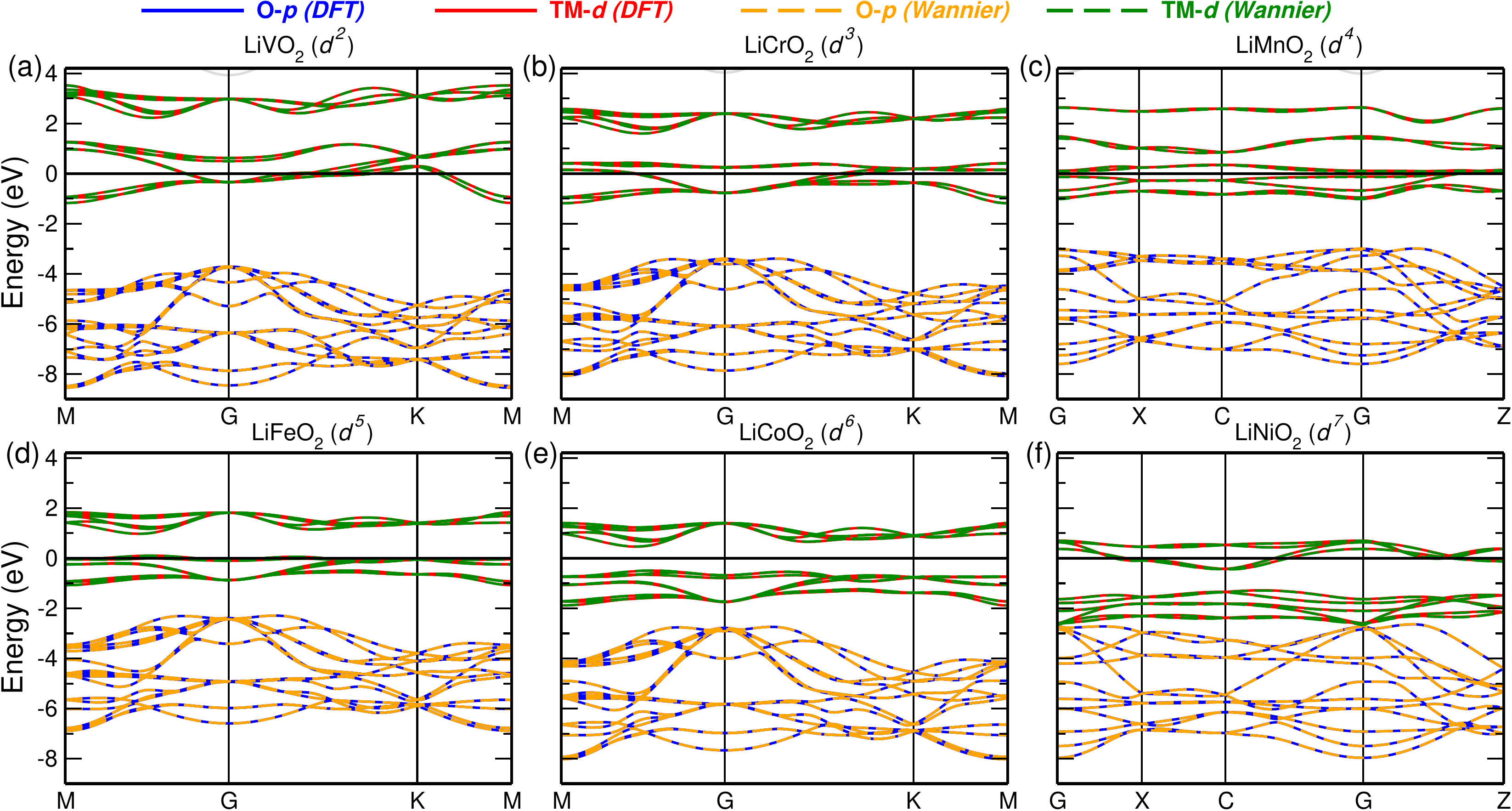}
\end{center}
\caption{The nonmagnetic DFT band structures of Li$M$O$_{2}$, along with their Wannier-projected bands. The solid and dashed lines represent the $Ab~initio$ and Wannier-projected bands, respectively.}
\label{fig:lmo-bs-1x}
\end{figure} 

\section{Supplementary Note 3}
We have considered three transition metal oxide perovskites, Sr$M$O$_{3}$ ($M$ = Mn, Fe, and Co), to study the interaction parameters for systems that exhibit the entangled / hybridized correlated space. The corresponding partial density of states (PDOS), shown in Fig.~\ref{fig:pdos-smo-1x}, and the orbital-resolved band structures, shown in Fig.~\ref{fig:orb-resolv-smo}, clearly demonstrate the strong $pd$-hybridization as electron filling increases.\\

\renewcommand{\figurename}{Fig.}
\renewcommand{\thefigure}{S\arabic{figure}}
\begin{figure}[h]
\begin{center}
\includegraphics[clip=true,scale=0.26]{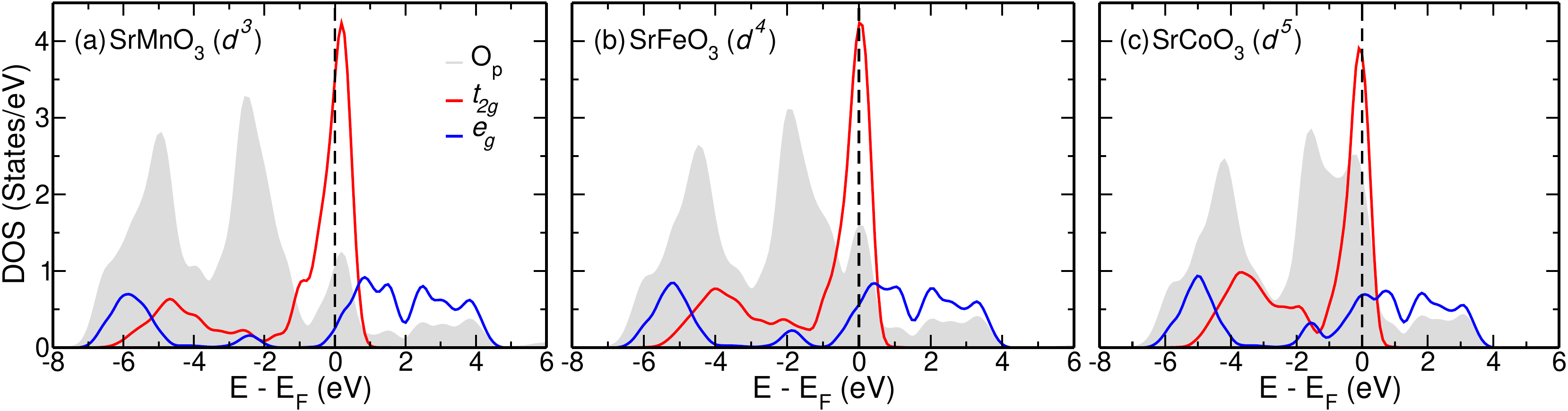}
\end{center}
\caption{Partial density of states (PDOS) of Sr$M$O$_{3}$ ($M$ = Mn, Fe, and Co). The Fermi level (E$_{F}$) is set to 0 eV.}
\label{fig:pdos-smo-1x}
\end{figure} 

\renewcommand{\figurename}{Fig.}
\renewcommand{\thefigure}{S\arabic{figure}}
\begin{figure}[h]
\begin{center}
\includegraphics[clip=true,scale=0.6]{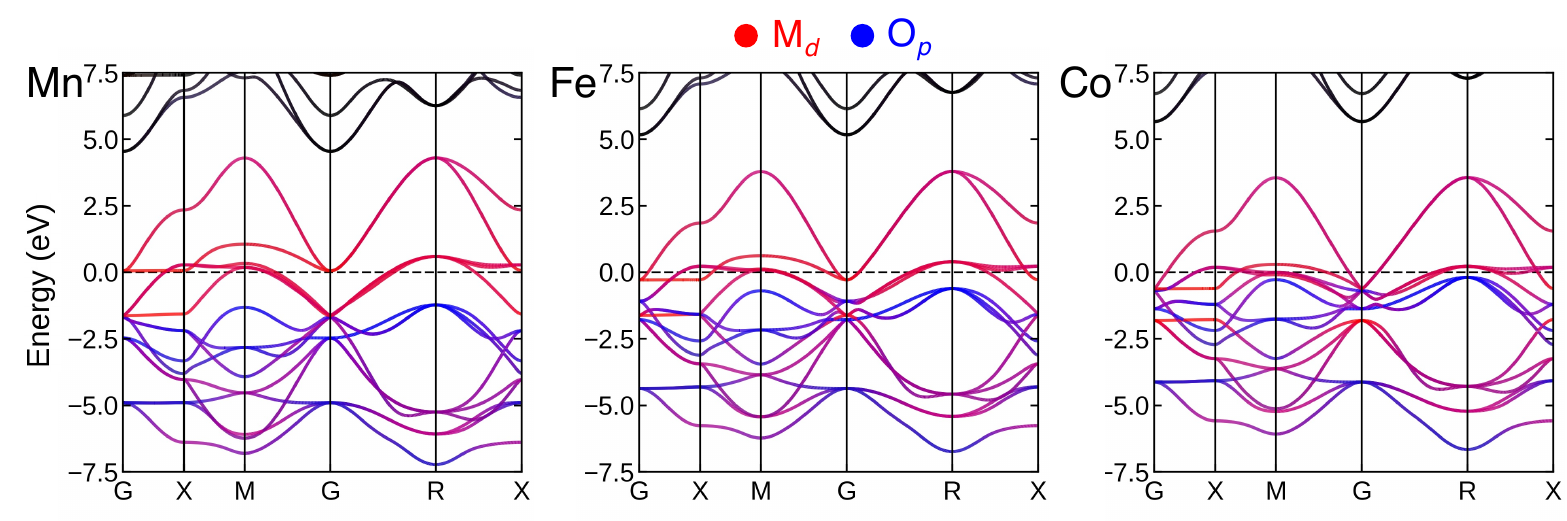}
\end{center}
\caption{Orbital-resolved band structures of Sr$M$O$_{3}$ ($M$ = Mn, Fe, and Co), with TM-$d$ and O-$p$ bands projected in red and blue, respectively.}
\label{fig:orb-resolv-smo}
\end{figure} 


Five sets of Wannier basis are constructed to study the screened interaction parameters using different cRPA methods, namely the projector method, the weighted method, and the disentanglement method. The corresponding projected Wannier bands, along with the calculated DFT band structures of Sr$M$O$_{3}$, are shown in Fig.~\ref{fig:smo-bs-SI}. The outer and frozen (inner) energy windows used to construct the various Wannier basis are provided in Table.~\ref{table:energy-windows-si} and marked in the band structures shown in Fig.~\ref{fig:smo-bs-SI}. Before discussing how the chosen different energy windows and basis affect the unscreened and screened parameters, we briefly outline our choices to obtain frontier and localized basis.
\begin{enumerate}[(i)]
    \item Frontier Orbitals 1 (FO1): A $d$-$d$ model with an outer energy window limited to $d$ bands only.
    \item Frontier Orbitals 2 (FO2): Same as FO1, but with the frozen window.
    \item Frontier Orbitals 3 (FO3): The outer energy window comprising the TM-$d$ and O-$p$ bands with frozen window. Note that only the outer energy window is extended to $dp$-energy range, but the O-$p$ bands are not included in constructing the Wannier functions.
    \item Atomic Orbitals 1 (AO1): A $d$-$d$ model but the outer energy window comprising both TM-$d$ and O-$p$ bands. Wannier functions are constructed only for the $d$-bands, without using a frozen window.
    \item Atomic Orbitals 2 (AO2): A hybrid $d$-$dp$ model. The outer energy window includes both TM-$d$ and O-$p$ bands and Wannier functions are constructed for all TM-$d$ and O-$p$ bands.  The Hubbard $U$ values are calculated for the $d$ manifold only.
\end{enumerate}

Here, we note that the energy window must be wide enough to encompass the bands of interest but not too broad to include additional bands beyond those of interest. Otherwise, extending the energy window too broadly can lead to poorly reproduced band dispersion. Although the broad energy window enhances the localization of Wannier functions, the resulting Wannier bands may become flat and narrow, with poor dispersion that does not match the DFT bands \cite{souza2001maximally}. In our study, this issue arises when all relevant bands within the outer energy window are not included to construct the Wannier functions, as seen in the case of AO1 basis.\\

\renewcommand{\figurename}{Fig.}
\renewcommand{\thefigure}{S\arabic{figure}}
\begin{figure}[h]
\begin{center}
\includegraphics[clip=true,scale=0.3]{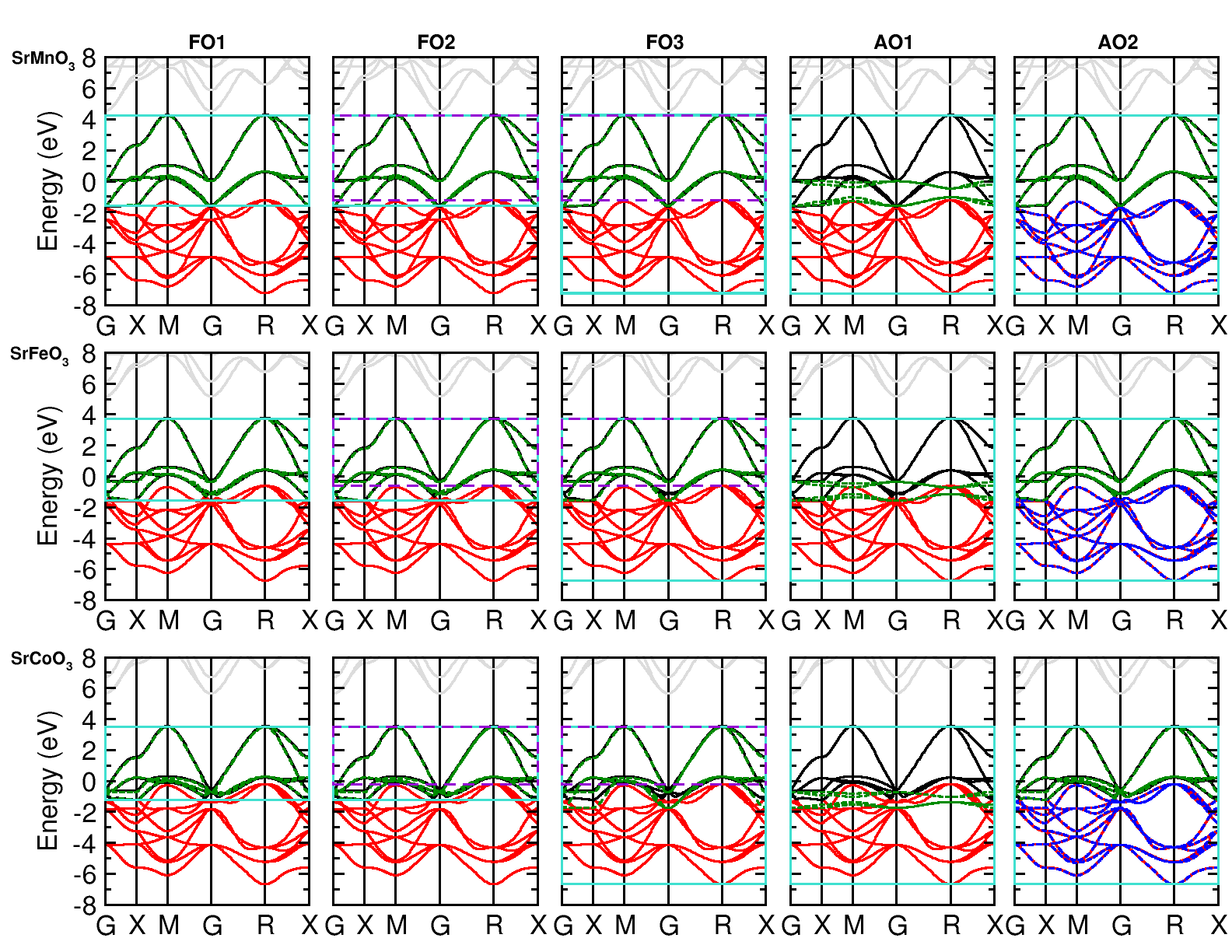}
\end{center}
\caption{The nonmagnetic DFT band structures along with their Wannier-projected bands for a simple cubic Sr$M$O$_{3}$. The solid black (red) and dashed green (blue) lines represent the $Ab~initio$ and Wannier-projected bands of TM-$d$ (O-$p$), respectively. The outer (frozen) energy windows are represented by solid turquoise (dashed purple) lines.}
\label{fig:smo-bs-SI}
\end{figure} 

\renewcommand{\figurename}{Fig.}
\renewcommand{\thefigure}{S\arabic{figure}}
\begin{figure}[h]
\begin{center}
 \includegraphics[clip=true,scale=0.5]{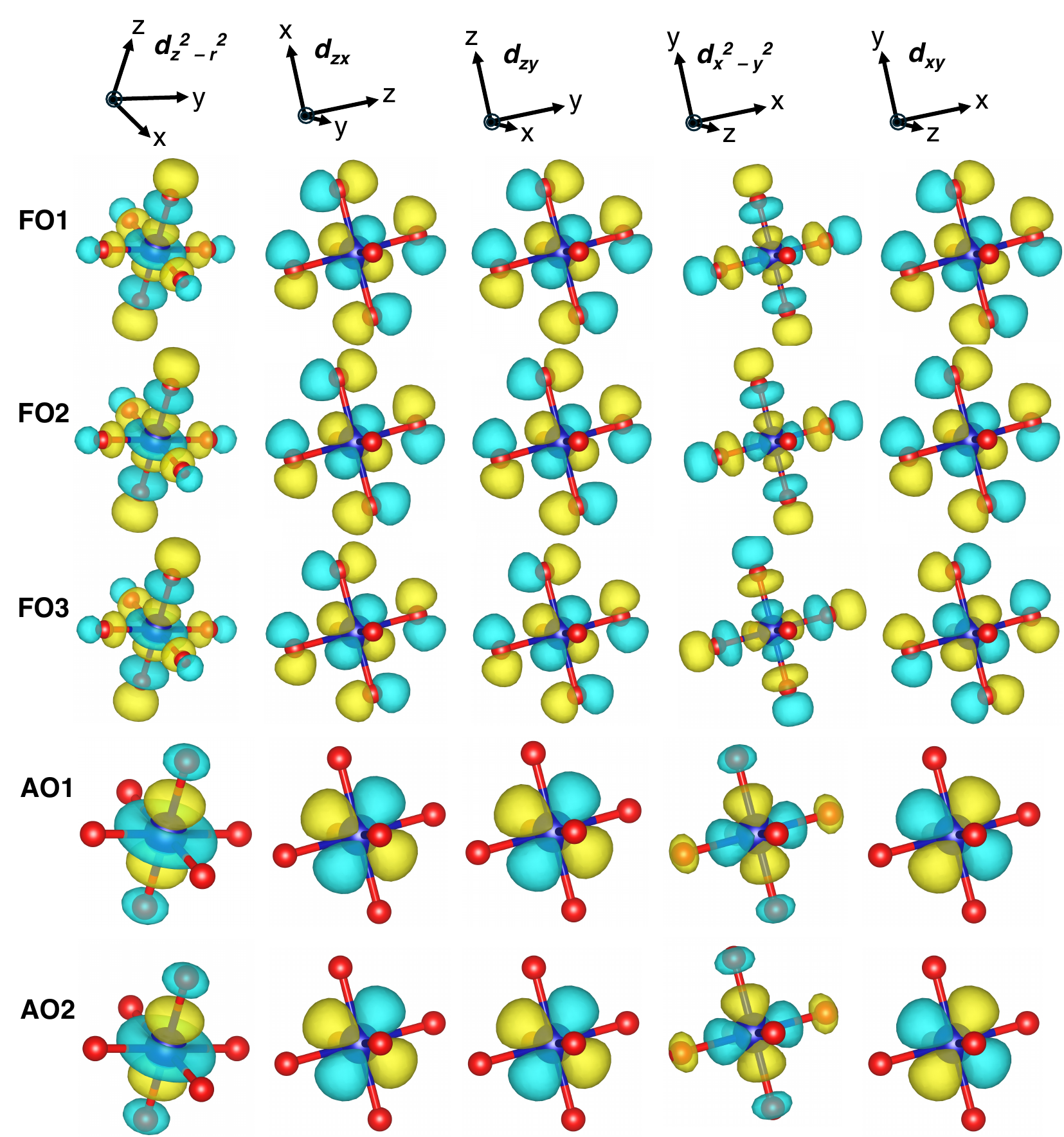}
\end{center}
\caption{The nonmagnetic DFT band structures along with their Wannier-projected bands for a simple cubic Sr$M$O$_{3}$. The solid black (red) and dashed green (blue) lines represent the $Ab~initio$ and Wannier-projected bands of TM-$d$ (O-$p$), respectively. The outer (frozen) energy windows are represented by solid turquoise (dashed purple) lines.}
\label{fig:wan-orbs-co}
\end{figure} 

\begin{table}[ht]
\begin{center}
\renewcommand{\tablename}{Table.}
\renewcommand{\thetable}{S\arabic{table}}
    \caption{The outer and frozen (inner) energy window ${\mathbb{W}}$ (in eV) and spread of Wannier functions (in \AA$^{2}$/orbital) for different basis for Sr$M$O$_{3}$ ($M$ = Mn, Fe, and Co).}
    \renewcommand{\arraystretch}{2}
    \begin{tabular} {c | c | c | c | c | c | c}
    \hline
    \hline
    ~ & ~ & FO1 & FO2 & FO3 & AO1 & AO2 \\
    \hline
    ~ & outer-${\mathbb{W}}$ & [-1.60,4.26] & [-1.60,4.26] & [-7.24,4.26] & [-7.24,4.26] & [-7.24,4.26] \\
    Mn & frozen-${\mathbb{W}}$ & - & [-1.23,4.26] & [-1.23,4.26] & - & - \\
    ~ & spread & 2.35 & 2.35 & 2.35 & 0.53 & 0.54 \\
    \hline
    ~ & outer-${\mathbb{W}}$ & [-1.54,3.74] & [-1.54,3.74] & [-6.75,3.74] & [-6.75,3.74] & [-6.75,3.74] \\
    Fe & frozen-${\mathbb{W}}$ & - & [-0.61,3.74] & [-0.61,3.74] & - & - \\
    ~ & spread & 5.74 & 5.00 & 2.63 & 0.48 & 0.48 \\
    \hline
    ~ & outer-${\mathbb{W}}$ & [-1.21,3.50] & [-1.21,3.50] & [-1.21,3.50] & [-6.67,3.50] & [-6.67,3.50] \\
    Co & frozen-${\mathbb{W}}$ & - & [-0.19,3.50] & [-0.19,3.50] & - & - \\
    ~ & spread & 5.36 & 5.69 & 3.03 & 0.44 & 0.44 \\
    \hline
    \hline
    \end{tabular}
    \label{table:energy-windows-si}
\end{center}
\end{table}

$Unscreened~Coulomb~interaction~U^{bare}$: In the case of the frontier basis (FOs), despite the localization of the $d$-orbitals, $U^{bare}$ values decrease with increasing electron occupation (see Table.~\ref{tab:bare_hubbard}), due to the inadequacy of the Wannier basis in capturing $pd$-hybridization within FOs. We now explore how this effect appears in each material.\\

For SrMnO$_{3}$, the spread and $U^{bare}$ are the same across all three sets of FOs, which is expected due to minimal entanglement between Mn-$d$ and O-$p$ bands. In the case of SrFeO$_{3}$, the $d$-bands outside the frozen energy window tends to localize, which is even more pronounced with FO3 basis at the $\Gamma$ point. As a result, $U^{bare}$ increases from FO1 to FO2, and further increases with FO3 basis. For SrCoO$_{3}$, the increased weight of the $d$-bands outside the frozen energy window, along with spatial constraints, limits their extension - unlike the behavior observed in SrFeO$_{3}$ with FO3 basis. This causes the $d$-bands to align with the DFT bands, albeit at the cost of an increased Wannier spread. However, the large outer energy window facilitates the localization of the $d$-bands, which is pronounced in the $d_{xy}$ Wannier orbital, where the signs of node centered on Co site change, consistent with AO behavior (see Fig.~\ref{fig:wan-orbs-co}). Thus, $U^{bare}$ decreases from FO1 to FO2, and then reverses for FO3 basis.\\

In the case of the localized basis (AO), $U^{bare}$ values increase with increasing electron occupation (see Table.~\ref{table:energy-windows-si}), which is reflected in the decrease in spread of Wannier functions (see Table.~\ref{tab:bare_hubbard}). The $U^{bare}$ values obtained from the localized basis are the same for both AO1 and AO2, due to the well-localized nature of the orbitals (see Fig.~\ref{fig:wan-orbs-co}).\\

The $J^{bare}$ follow a similar trend to $U^{bare}$ for all three FO basis sets and two AO basis sets (see Table.~\ref{tab:bare_hubbard}).\\

\begin{table}[!htb]
\begin{center}
\renewcommand{\tablename}{Table.}
\renewcommand{\thetable}{S\arabic{table}}
    \caption{The Unscreened Hubbard parameters ($U^{bare}$ and $J^{bare}$) different basis sets for Sr$M$O$_{3}$ ($M$ = Mn, Fe, and Co).}
    \renewcommand{\arraystretch}{2}
    \begin{tabular} {c | c | c | c | c | c | c}
    \hline
    \hline
     ~ & \multicolumn{3}{c|}{$U^{bare}$ (in eV)} & \multicolumn{3}{c}{$J^{bare}$ (in eV)} \\
    \cline{2-7}
    ~ & Mn & Fe & Co & Mn & Fe & Co \\
    \hline
    FO1 & 16.54 & 15.03 & 13.78 & 0.49 & 0.42 & 0.37 \\
    FO2 & 16.53 & 15.04 & 13.52 & 0.49 & 0.42 & 0.36 \\
    FO3 & 16.53 & 16.21 & 16.03 & 0.49 & 0.47 & 0.45 \\
    AO1 & 23.59 & 24.88 & 26.19 & 0.80 & 0.83 & 0.87 \\
    AO2 & 23.65 & 24.94 & 26.24 & 0.80 & 0.83 & 0.87 \\
    \hline
    \hline
    \end{tabular}
    \label{tab:bare_hubbard}
\end{center}
\end{table}

$Screened~Coulomb~interaction~(U)$: In the FO basis, the calculated $U$ values using the disentanglement method are larger than those obtained from the projector and weighted methods. This increase is notable despite the slight overlap between the Mn-$d$ and underlying O-$p$ bands in SrMnO$_{3}$. 
This discrepancy arises because the disentanglement method strongly depends on the chosen energy window and does not consider the hybridization between correlated and rest spaces, which affects the band structure. In contrast, the projector and weighted methods work on the unaltered band structure, which explains the divergence in $U$ values obtained by the disentanglement method compared to the other two methods. Fig.~\ref{fig:methods-smo} illustrates that $U$ values obtained using the disentanglement method are identical for the FO1 and FO2 basis but decrease for the FO3 basis as the energy window increases.\\

For SrMnO$_{3}$, the projector and weighted methods yield identical $U$ values for all FO basis sets due to the minimal entanglement between the Mn-$d$ and O-$p$ bands. In fact, this minimal entanglement also results the similar $U$ values for the AO basis sets in the projector method. For SrFeO$_{3}$ and SrCoO$_{3}$, the increased entanglement / hybridization leads to a slight increase in the $U$ values obtained from the FO3 basis compared to FO1 and FO2 basis sets.\\

In the AO basis, the $U$ values calculated from AO1 basis are smaller than those from the AO2 basis, which is particularly noticeable when using the projector method. The weighted method predicts smaller $U$ values compared to the projector method, as it neglects the screening effects within the correlated space. Again, the disentanglement method results in larger $U$ values.\\

\renewcommand{\figurename}{Fig.}
\renewcommand{\thefigure}{S\arabic{figure}}
\begin{figure}[h]
\begin{center}
\includegraphics[clip=true,scale=0.45]{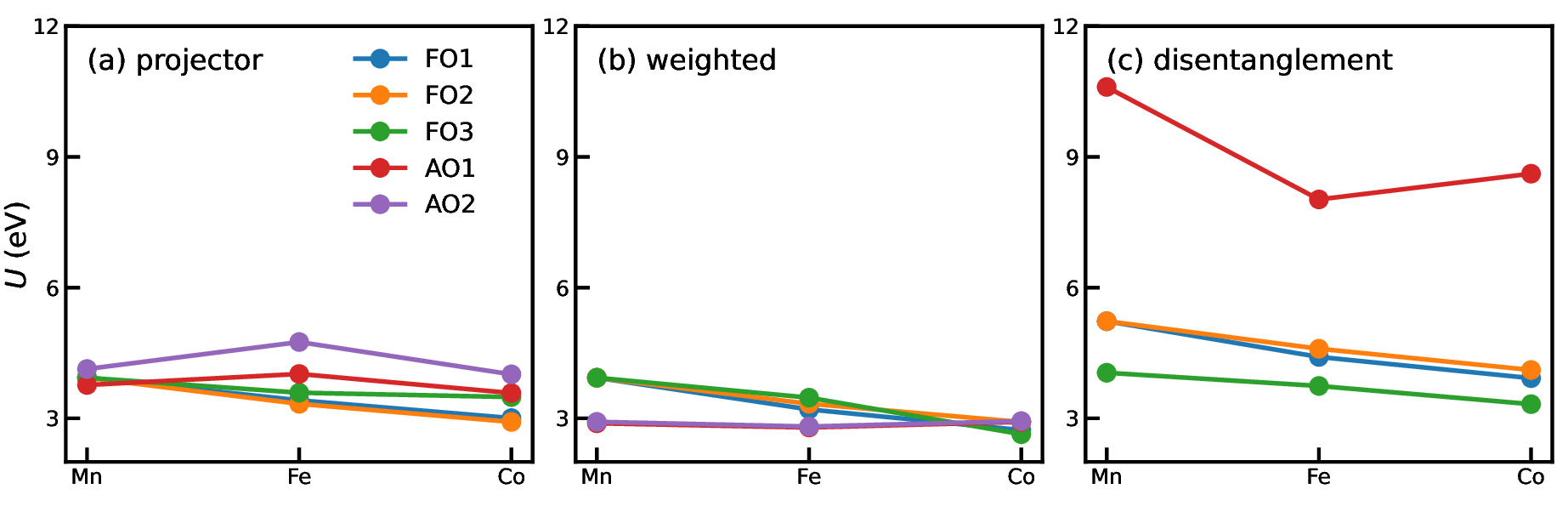}
\end{center}
\caption{Comparison of Hubbard $U$ values calculated for different basis across Sr$M$O$_{3}$ systems.}
\label{fig:methods-smo}
\end{figure}

\bibliography{si}